\documentclass[11pt]{article}
\pdfoutput=1
\usepackage{jcappub,natbib}
\input{colordvi.tex}

\usepackage{amsmath,amssymb,graphicx}

\usepackage{subfigure}

\usepackage{mathtools}

\def\ga{\mathrel{\raise.3ex\hbox{$>$\kern-.75em\lower1ex\hbox{$\sim$}}}}
\def\la{\mathrel{\raise.3ex\hbox{$<$\kern-.75em\lower1ex\hbox{$\sim$}}}}

\def\Msun{M_\odot}

\newcommand{\be}{\begin{eqnarray}}
\newcommand{\en}{\end{eqnarray}}
\newcommand{\nn}{\nonumber\\}

\newcommand{\braket}[1]{\langle #1 \rangle}

\newcommand{\mPBH}{m_{\rm{PBH}}}
\newcommand{\dd}{\mathrm{d}}
\newcommand{\rr}{\mathrm}

\title{Detecting the Stochastic Gravitational Wave Background from Primordial Black Hole Formation}

\author[a,b]{S\'ebastien Clesse}
\author[c,d]{Juan Garc\'ia-Bellido,}
\author[a]{Stefano Orani,}

\affiliation[a]{Cosmology, Universe and Relativity at Louvain (CURL), Institut de Recherche en Mathematique et Physique (IRMP), Louvain University, 2 Chemin du Cyclotron, 1348 Louvain-la-Neuve, Belgium}
\affiliation[b]{Namur Center of Complex Systems (naXys), Department of Mathematics, University of Namur, Rempart de la Vierge 8, 5000 Namur, Belgium}
\affiliation[c]{Instituto de F\'isica Te\'orica UAM-CSIC, Universidad Auton\'oma de Madrid, Cantoblanco, 28049  Madrid, Spain}
\affiliation[d]{CERN, Theoretical Physics Department, 1211 Geneva, Switzerland}

\abstract{
Primordial Black Holes (PBH) from peaks in the curvature power spectrum could constitute today an important fraction of the Dark Matter in the Universe. At horizon reentry, during the radiation era, order one fluctuations collapse gravitationally to form black holes and, at the same time, generate a stochastic background of gravitational waves coming from second order anisotropic stresses in matter.    We study the amplitude and shape of this background for several phenomenological models of the curvature power spectrum that can be embedded in waterfall hybrid inflation, axion, domain wall, and boosts of PBH formation at the QCD transition.  
For a broad peak or a nearly scale invariant spectrum, this stochastic background is generically enhanced by about one order of magnitude, compared to a sharp feature.  As a result, stellar-mass PBH from Gaussian fluctuations with a wide mass distribution are already in strong tension with the limits from Pulsar Timing Arrays, if they constitute a non negligible fraction of the Dark Matter.  But this result is mitigated by the uncertainties on the curvature threshold leading to PBH formation.   LISA will have the sensitivity to detect or rule out light PBH down to $10^{-14} M_{\odot}$.    Upcoming runs of LIGO/Virgo and future interferometers such as the Einstein Telescope will increase the frequency lever arm to constrain PBH from the QCD transition.  Ultimately, the future SKA Pulsar Timing Arrays could probe the existence of even a single stellar-mass PBH in our Observable Universe.
}

\begin{document}

\begin{flushright}  IFT--UAM/CSIC--17--119,  CERN--TH--2017--259 \end{flushright}

\maketitle
\flushbottom

%\begin{itemize}
%\item Introduction. We describe the state of the art in GW and PBH
%\item Models. We describe the different models of PBH production and their range of masses and shapes of power spectra 
%\item Computation of second order GW background. Here is the main body of the text.
%\item Results for each model. Evaluate the background for Gaussian, broken-power-law and delta-function P(k).
%\item Expectations from future detectors: ground and space interferometers, as well as PTA.
%\item Conclusions
%\item Figures: \\ 
%         Fig.1 - H(a) and horizon crossing. Three oblique lines (2 red and 1 blue) \\  
%         Fig.2 - Curvature power spectrum in k (1/Mpc) and M (Msun) for CGB model. Three vertical lines corresp. to Fig.1\\
%         Fig.3 - Same for axion model with narrow spectrum or delta function \\ 
%         Fig.4 - Same for broken power law \\ 
%         Fig.5 - Spectrum of GW today in $\OGW$ for $\Pi^2=100$ as a function of f(Hz). Show curves for different masses. \\
%         Fig.6 - Spectra for each model compared with LIGO, LISA and PTA sensitivites.

%\end{itemize}  

\section{Introduction}\label{sec:intro}

Since the detection by Advanced LIGO/VIRGO of Gravitational Waves from the merging of massive black holes (BH)~\cite{Abbott:2016blz,Abbott:2016nmj,TheLIGOScientific:2016pea,Abbott:2016bqf,Abbott:2017vtc,Abbott:2017gyy,Abbott:2017oio}, there has been a strong revival of interest for Primordial Black Holes (PBH)~\cite{Hawking:1971ei,Carr:1974nx,1975Natur.253..251C,Carr:1975qj,GarciaBellido:1996qt} as a Dark Matter (DM) candidate~\cite{Clesse:2015wea,Bird:2016dcv,Clesse:2016vqa,Sasaki:2016jop}.  Several observations motivate this scenario~\cite{Garcia-Bellido:2017fdg,Clesse:2017bsw}:  in particular, the merging rates inferred by LIGO that are consistent with the ones expected for PBH abundances comparable to DM; the low effective spin of BH progenitors, one of them being likely anti-aligned with the orbital momentum, which suggests a capture process; recently detected microlensing events of stars in M31, the Milky Way bulge and distant quasars; the lack of detections of ultra-faint dwarf galaxies (or their stellar clusters) below the critical dynamical heating radius; the spatial coherence between the source-subtracted soft X-ray and cosmic infrared backgrounds~\cite{Kashlinsky:2016sdv,Cappelluti:2017nww}; the existence of billions of super-massive black holes at high redshifts that could have grown from massive PBH seeds in the tail of the mass distribution. Note that PBH as DM could also help resolve the long-standing core-cusp, too-big-to-fail and dwarf satellite problems of LSS, thanks to the gravitational scattering of PBH in the halo cores as well as an early and rapid accretion phase in dense halos~\cite{Clesse:2017bsw}.

In particular, a PBH-DM scenario with an extended mass distribution centered in the range $[1-10]\, \Msun$ could perfectly explain the detected LIGO events and pass all the present observational constraints~\cite{Khlopov:2008qy,Carr:2009jm,Carr:2016drx,Bellomo:2017zsr,Carr:2017jsz}, if the uncertainties on microlensing limits~\cite{Hawkins:2011qz,Hawkins:2015uja,Green:2017qoa,Calcino:2018mwh} or the possible clustering of PBH~\cite{Garcia-Bellido:2017xvr} are taken into account.  Notice, however, that it is still unclear and debated whether numerous BH binaries would form soon after PBH generation, and boost the merging rate today above the LIGO limits~\cite{   TheLIGOScientific:2016wyq,Mandic:2016lcn,Clesse:2016ajp,Kocsis:2017yty,Raidal:2017mfl,Ali-Haimoud:2017rtz,Bringmann:2018mxj,Ali-Haimoud:2018dau,Belotsky:2018wph,Raidal:2018bbj}, in particular if PBH are initially clustered and have an extended mass distribution.

A nice feature of the scenario is that within the next decade, it will be possible to prove or rule out massive PBH in the range $[0.1-10] \Msun$ constituting all of the DM, with a set of independent but complementary observations, from widely different scales and epochs in the history of the Universe~\cite{Garcia-Bellido:2017fdg,Clesse:2017bsw}.  Among others, it has been proposed to probe PBH and their mass distribution by using the BH merging rate distribution from GW detectors, the stochastic GW background from PBH binaries~\cite{Mandic:2016lcn,Clesse:2016ajp,Kovetz:2017rvv,Raidal:2017mfl} or close PBH encounters~\cite{Garcia-Bellido:2017qal,Garcia-Bellido:2017knh}, by detecting many new ultra-faint dwarf galaxies (UFDG) and their luminous density profile~\cite{Brandt:2016aco,Green:2016xgy,Li:2016utv}, the correlation of radio and high-energy sources towards the galactic center~\cite{Gaggero:2016dpq,Bartels:2015aea,Lee:2015fea}, wide binaries in the galactic halo~\cite{Quinn:2009zg}, microlensing events~\cite{Calcino:2018mwh}, the imprints of early PBH accretion on the cosmic microwave background anisotropies and spectral distortions~\cite{Ricotti:2007au,Ali-Haimoud:2016mbv,Poulin:2017bwe}, and on the 21cm signal~\cite{Tashiro:2012qe,Gong:2017sie,Gong:2018sos,Clark:2018ghm,Hektor:2018qqw}, the correlation between the cosmic infrared background and the soft X-ray background~\cite{Kashlinsky:2016sdv,Cappelluti:2017nww}, etc.  

Together with observational probes, on the theory side, several models for massive PBH formation have been proposed recently in the context of inflation, including mild-waterfall hybrid inflation~\cite{Clesse:2015wea}, axion-gauge inflation~\cite{Linde:2012bt,Bugaev:2013fya,Erfani:2015rqv,Cheng:2016qzb,Garcia-Bellido:2016dkw,Garcia-Bellido:2017aan}, critical Higgs inflation~\cite{Garcia-Bellido:2017mdw,Ezquiaga:2017fvi,Motohashi:2017kbs,Bezrukov:2017dyv}, inflection point~\cite{Germani:2017bcs} etc.   A common feature of all these models is a peak in the power spectrum of curvature fluctuations.  Another interesting possibility is to produce a PBH abundance peaked on the solar-mass scale from a nearly scale invariant power spectrum, thanks to the sound speed reduction during the QCD phase transition~\cite{Jedamzik:1996mr,Cardall:1998ne,Byrnes:2018clq}.  Therefore, if the existence of PBH were confirmed, it would be possible to constrain and distinguish these different models and formation mechanisms.  Future LSS surveys, CMB spectral distortions~\cite{PIXIE,PRISM} and limits on the abundance of ultra-compact mini-halos~\cite{Bringmann:2011ut,Emami:2017fiy} will set limits on scales where power enhancement could be initiated.  But for the moment there are no observations able to probe the entire shape of the power spectrum peak~\cite{Garcia-Bellido:2017aan}.

In this paper, we study a potentially detectable effect of PBH formation from peaks in the curvature power spectrum.  At horizon reentry, during the radiation era, these large fluctuations collapse gravitationally to form PBH and, at the same time, they generate a stochastic background of gravitational waves from the second-order anisotropic stress-tensor of matter fluctuations~\cite{Ananda:2006af,Baumann:2007zm,Inomata:2016rbd,Nakama:2016gzw,Orlofsky:2016vbd,Garcia-Bellido:2017aan,Gong:2017qlj,Inomata:2018epa}.   Compared to other recent works, focusing on a monochromatic PBH mass~\cite{Inomata:2016rbd,Orlofsky:2016vbd}, on the effect of non-Gaussianity~\cite{Nakama:2016gzw} or on specific models~\cite{Orlofsky:2016vbd,Gong:2017qlj,Garcia-Bellido:2017aan}, our analysis is extended to more realistic wide-mass distributions of PBH.   The amplitude and shape of this new background is calculated here for four different phenomenological models for the primordial power spectrum of curvature fluctuations, able to generate massive PBH as the DM:  a broad Gaussian peak, a very narrow peak behaving in the limit like a $\delta$-function, a broken power-law and a nearly scale-invariant spectrum.   For a Gaussian peak, our result confirm independently the ones of~\cite{Inomata:2018epa} that were released during the finalization of our paper.  These typical spectral shapes reproduce the predictions of different theoretical models such as waterfall hybrid inflation, axion-gauge inflation, domain wall, and QCD phase transition models.   We identify universal features and frequency ranges of this GW background that would allow one to distinguish it from other backgrounds with future ground and space interferometers, as well as future Pulsar Timing Arrays (PTA).   It is also shown that the present limits on PTA already exclude  part of the parameter space of these models.   Finally, we will emphasize that future PTA and the LISA mission can potentially exclude the existence of even a single PBH in our observable Universe, if these PBH are produced from Gaussian fluctuations. 

The paper is organized as follows:  The formalism of PBH formation from peaks in the curvature spectrum is reviewed in Section~\ref{sec:PBHform}.  Then, the considered models and the corresponding curvature power spectra are presented in Section~\ref{sec:models}.  In Section~\ref{sec:gw}, we compute the expected stochastic GW spectra from the PBH formation and we discuss their detectability in Section~\ref{sec:obs}.  We conclude and discuss some perspectives of this work in Section~\ref{sec:conclusions}.

\section{PBH formation from peaks in the curvature power spectrum}  
\label{sec:PBHform}

PBH can form during the radiation era due to the gravitational collapse of $\mathcal O(1)$ curvature fluctuations, when they reenter inside the Hubble horizon.   We focus here on  Gaussian fluctuations, produced during inflation.   We adopt a model independent approach and consider typical shapes of the small-scale power spectrum peak, at the origin of PBH formation.   Some level of non-Gaussianity could ease the formation of PBH, as shown e.g. in~\cite{Franciolini:2018vbk,Young:2015kda,Ezquiaga:2018gbw,Byrnes:2012yx,PinaAvelino:2005rm}.   But here we assumed that the condition derived in~\cite{Franciolini:2018vbk} (see Eq. 3.6) for non-Gaussianity to not alter significantly the PBH abundance is satisfied.   Our results could nevertheless be extended to a scenario where non-Gaussian fluctuations are important.  Typically this would lower the limit on the peak amplitude, leading to a lower amplitude of the stochastic background from PBH formation, as discussed later.

There is a one-to-one correspondance between the PBH mass $\mPBH$, that is roughly the mass in the collapsing Hubble volume, the time $t_k$ and scale factor $a_k$ at formation in the radiation era, and the wavenumber $k$ associated to the size of the curvature fluctuation that had exit the Hubble radius $N_k$ e-folds before the end of inflation, so that
\be
\mPBH(t_k) = \frac{M_{\rm p}^2}{H_k} =  \frac{M_{\rm p}^2}{H_\rr{end}} \left( \frac{a_{ k}}{a_{\rm rh}}  \right)^2 \left( \frac{a_{\rm rh}}{a_{\rm end}} \right)^{3 (1+w_{\rm rh}) /2}
%= \frac{M_{\rm p}^2}{H_\rr{inf}} \rr e^{2 N_k}~,
\en
where $M_{\rm p}$ is the reduced Planck mass, $H_k$ and $a_{k} $ are the Hubble rate and scale factor at formation time $t_k$, $H_\rr{end}$ and $a_{\rm end}$ are the ones at the end of inflation.  Here $a_{\rm rh} $ denotes the scale factor at the end of the reheating era, which we assumed has an effective equation-of-state parameter $w_{\rm rh} $.   For simplicity, we will assume instantaneous reheating.   This way, $\mPBH(t_k) = M_{\rm p}^2/ H_{\rm end} \exp (2 N_k) $.    One can notice that the shape of the curvature power spectrum $\mathcal P_\zeta(N_k)$ alone does not determine a unique PBH mass.  For doing so, one needs to fix the energy scale of inflation, or use instead $\mathcal P_\zeta(k)$.    We assumed a standard $\Lambda$CDM cosmology with $H_0=70\,{\rm km/s/Mpc}$ and matter-radiation equality at $a_{\rm eq}=3.4\times10^{-4}$.

The fraction $\beta$ of the Universe collapsing into black holes of mass between $\mPBH$ and $ \mPBH + \dd  \ln  \mPBH $ is given by
\be  \label{eq:betaform}
\beta_{\rm form}(\mPBH)  \equiv \frac{1} {\rho_{\rm tot}} \frac{\dd \rho_{\rm PBH} (m) }{\dd \ln \mPBH}
 = 2 \int_{\zeta_{\rr c}}^{\infty} \frac{1}{\sqrt{2 \pi } \sigma} \rr e^{- \frac{\zeta^2}{2 \sigma^2} }\dd \zeta =  \rr{erfc} \left( \frac{\zeta_{\rm c}}{\sqrt{2} \sigma} \right)~,
\en
where $ \zeta_{\rr c} $ is the curvature fluctuation threshold above which gravitational collapse leads to the formation of a PBH.  In the limit $\zeta_{\rm c} \gg \sqrt 2 \sigma$, this gives
\be \label{eq:betaaprrox}
\beta_{\rm form}(\mPBH)  \simeq ~ \frac{\sqrt 2 \sigma}{\sqrt \pi \zeta_{\rm c}} {\rm e}^{-\frac{\zeta_{\rm c}^2}{2 \sigma^2}}.
\en
For a pure radiation fluid, numerical relativity simulations in spherical symmetry give a density fluctuation threshold value  $ \delta_{\rr c} = 0.453$~\cite{Musco:2004ak,Shibata:1999zs}, which corresponds to $ \zeta_{\rr c} = 1.02$ at the Hubble crossing scale. Except for the model of Sec.~\ref{ssec:QCDpeak} in which we take into account the effect of the QCD transition on the equation of state of the plasma, we have assumed this value throughout the paper.  Let us nevertheless notice that a three-zone model describing the PBH collapse has been used to obtain a lower value, $\zeta_{\rm c} = 0.086$~\cite{Harada:2013epa}.  The impact of the critical threshold on our results will be discussed in Section~\ref{sec:obs}.  Finally, $\sigma$ is the variance of curvature fluctuations and is related to their primordial power spectrum convolved with some window function $W(k,R)$, assumed to be a simple top-hat function centered on $k_{\rm PBH} $ and over which the power spectrum varies only linearly\footnote{see~\cite{ Polnarev:2006aa,Nakama:2013ica,Musco:2018rwt,Germani:2018jgr} for detailed discussions on the possible effect of the density profile and of the window function. }, 
\be
\sigma^2 = \int _0^\infty W^2(k,R) \mathcal P_\zeta (k) \dd \ln k \simeq \mathcal P_\zeta (k_{\rr PBH}).
\en
After PBH formation, $\beta $ grows linearly with the scale factor in the radiation-dominated era, from the relative dilution due to the universe expansion of matter and radiation densities, and assuming no mass accretion during the radiation era.   At matter-radiation equality, one has
\be
\beta_{\rm eq} (\mPBH) \approx \frac{a_{\rm eq}}{a_{\rm form}} \beta_{\rm form} (\mPBH)~.
\en
Typically, for solar-mass PBH, one  needs $\beta_{\rm form} \sim 10^{-9} $ at formation to get an abundance comparable to the one of Dark Matter.

\begin{figure}[!tb]
\centering
\includegraphics[width=0.90\textwidth]{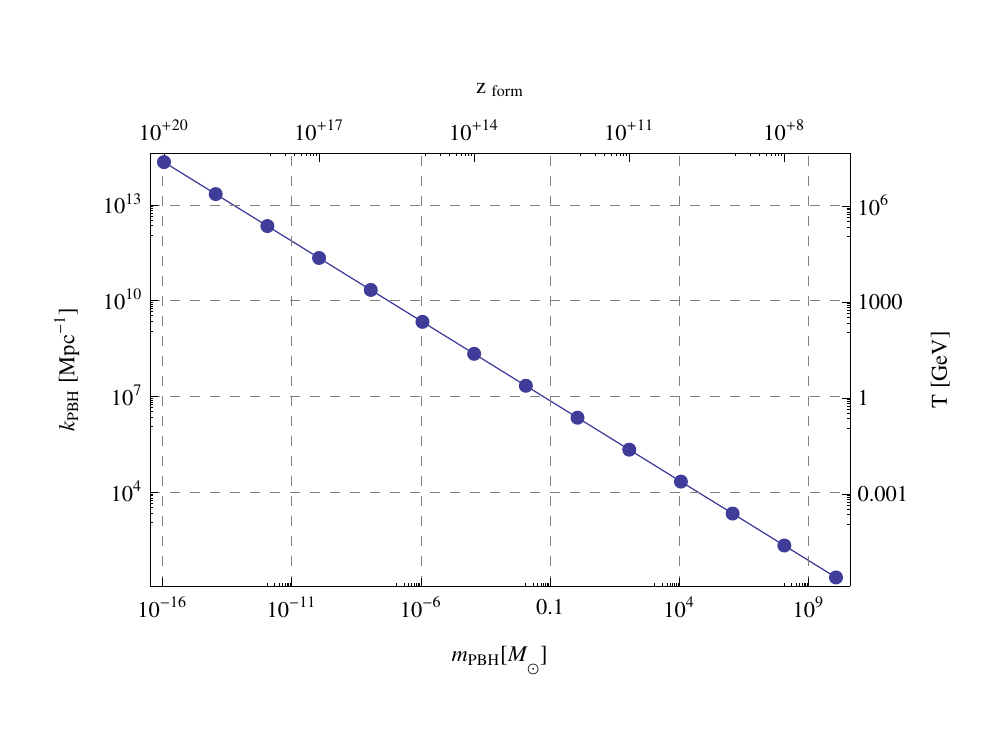}
\caption{Correspondance between PBH mass $\mPBH$, wavelength mode at the origin of PBH formation, redshift and temperature at formation time.  }
\label{fig:scales}
\end{figure}

\section{Models}
\label{sec:models}

%\begin{figure}[!tb]
%\centering
%\includegraphics[width=0.85\textwidth]{graphics/scales.pdf}
%\caption{Evolution of $H$ (black) and $k_*/a$ (dashed black), $k_{\rm p}\times10^{\pm 4.5}/a$ (red) and $k_{\rm p}/a$, for $k_*=0.05\,{\rm Mpc^{-1}}$ and $k_{\rm p}=1.23\times10^6\,{\rm Mpc^{-1}}$ versus $a/a_{0}$, where $a_{0}$ is the scale factor today. All the scales considered re-enter the horizon during radiation domination.}
%\label{fig:scales}
%\end{figure}

In this Section, we introduce the models of PBH formation that we consider and in Section~\ref{sec:gw} we compute the gravitational waves produced in these scenarios. 
The corresponding PBH mass distributions are represented on Fig.~\ref{fig:massdist}.

\subsection{Gaussian power spectrum}
\label{ssec:hi}

The first considered model is a Gaussian power spectrum, which arises rather generically in some versions of smooth-waterfall hybrid inflation~\cite{Clesse:2015wea}, as well as in some models of single field inflation with an inflection point~\cite{Germani:2017bcs}. The potential has a plateau for a brief period of the field evolution, before ending inflation, which induces a peak in the power spectrum at small scales. 
This process generates a characteristic peak in the spectrum of the primordial curvature perturbation $\mathcal{P}_\zeta$, shown in Fig.~\ref{fig:specs}. To a good approximation, it is given by
\be
\mathcal{P}_\zeta(N_k)\,\simeq\,A_s\left(\frac{k}{k_*}\right)^{n_s-1} + 
\mathcal{P_{\rr p}} \,
%\frac{\mathcal{P}_p}{\sqrt{2\pi}\,\sigma_{\rm p}}\,
\exp\left[-\frac{(N_k-N_p)^2}{2\sigma_{\rm p}^2}\right]\,.
\label{eq:phi}
\en 
%where $N_k=\ln(k/k_*)+N_*$. 
We assumed that the parameters  $\mathcal{P}_{\rm p}$ and $\sigma_{\rm p}$ describing the peak position and width, are essentially uncorrelated so that we can have any value for the width of the Gaussian in $N_k$, and any amplitude $\mathcal{P}_p$. We will choose here as characteristic of the Gaussian power spectrum a broad case with $\sigma_{\rm p} \sim 3$, and leave for section~\ref{ssec:delta} the case $\sigma \ll 1$, which at the limit corresponds to a delta function power spectrum.   

One may notice that the required amplitude of the broad peak to produce a significant fraction of dark matter in PBH only slightly depends on $\sigma_{\rm p}$.   The resulting PBH abundance is only sensitive to the top of the peak, due to the exponential dependence in $\beta_{\rm form}$, see Eq.~(\ref{eq:betaaprrox}), which makes quite generic the predictions for the maximal amplitude of the stochastic GW background associated to PBH formation.   For instance, for $\sigma_{\rm p} = 10$ and $k_{\rm p} = 2 \times 10^{6}\ {\rm Mpc}^{-1}$ (e.g. a peak centered on $1 M_\odot$) one needs $\mathcal P_{\rm p} \simeq 0.022 $ to get 100\% of DM made of PBH.  For a narrower peak with $\sigma_{\rm p} = 3$ (our benchmark case), one needs $\mathcal P_{\rm p} \simeq 0.025 $, only 10\% higher.

\subsection{Single sharp peak power spectrum}
\label{ssec:delta}

%\begin{figure}[!tb]
%\centering
%\includegraphics[width=0.85\textwidth]{graphics/Pdelta.pdf}
%\hfill
%\caption{}
%\label{fig:delta}
%\end{figure}

There are several models that present sharp peaks in the curvature power spectrum at a single scale, like the original model of hybrid inflation~\cite{GarciaBellido:1996qt,Lyth:2012yp}, or the more recent one of axion inflation with couplings to gauge fields~\cite{Linde:2012bt,Bugaev:2013fya,Erfani:2015rqv,Cheng:2016qzb,Garcia-Bellido:2016dkw,Garcia-Bellido:2017aan}. In both cases, the backreaction of explosive particle production at a given moment before the end of inflation produces a sharp peak in the power spectrum that has the form:
\be
\mathcal{P}_\zeta (k)\,=\, A_s\left(\frac{k}{k_*}\right)^{n_s-1} + \mathcal{P}_p \,\delta(k-k_{\rm p})\,,
\label{eq:delta}
\en
where for analytical treatments we have approximated the sharp peak with a delta function, but the differences for GW production at reentry are negligible.   For the numerical computation of the GW spectrum,  we consider the limit case of a thin Gaussian peak, shown on Fig.~\ref{fig:specs}, with $\sigma_{\rm p} =  0.1$ and $\mathcal P _{\rm p}  = 0.0298 $, such that $f_{\rm PBH}^{\rm tot} \equiv \Omega_{\rm PBH} / \Omega_{\rm DM} =1 $ today.

\subsection{Broken power law power spectrum}
\label{ssec:bpl}

%\begin{figure}[!tb]
%\centering
%\includegraphics[width=0.85\textwidth]{graphics/Pzbr.pdf}
%\hfill
%\caption{Spectrum of primordial curvature perturbations at the end of inflation $\mathcal{P}_{\zeta}$ as a function of $N_k$ for model (\ref{eq:pl}). The parameters are $P_p=0.1$, $m=3$ and $n=0.5$. The energy scale of inflation is fixed to $\Lambda=5.7\times10^{-43}$ by requiring that the PBH corresponding to the peak scale $k_{\rm p}=1.23\times10^6\,{\rm Mpc^{-1}}$ have a mass $M_{\rm PBH}=M_{\odot}$, consistently with figs.~\ref{fig:scales} and \ref{fig:specs}. The red vertical lines illustrate the window of modes considered for the evaluation of gravitational waves from PBH formation. In contrast to model \ref{eq:pot}, the broken power law spectrum is very asymmetric, meaning that the ranged of scales considered is biased towards the ultraviolet. As in the previous figures, the black dashed line indicates the pivot scale $k_*=0.05\,{\rm Mpc^{-1}}$.}
%\label{fig:pbr}
%\end{figure}

\begin{figure}[!tb] 
\centering
\includegraphics[width=0.77\textwidth]{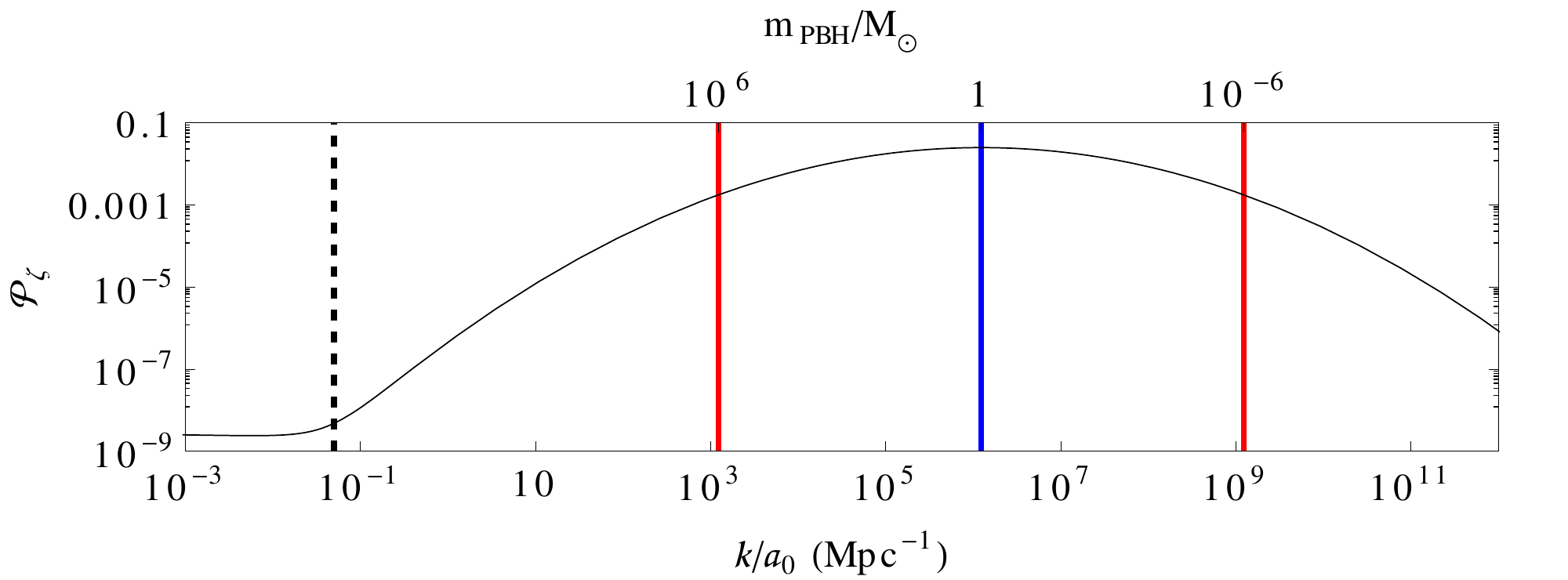}
\includegraphics[width=0.77\textwidth]{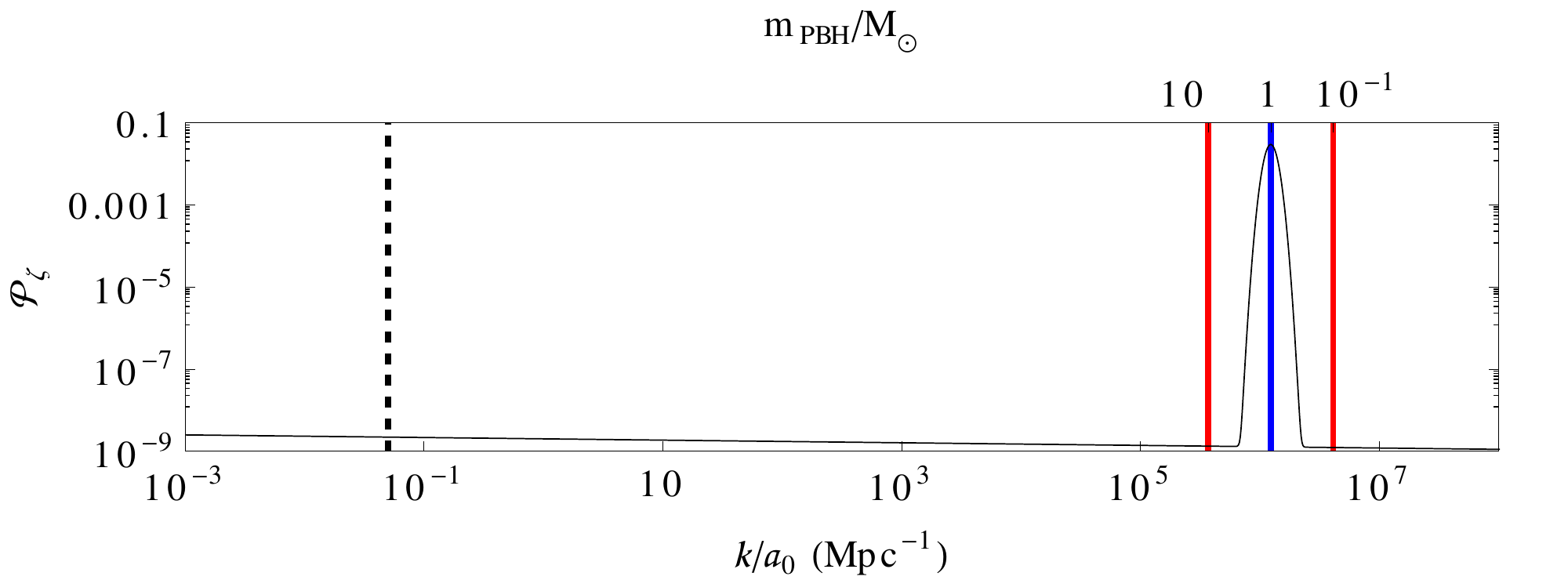}
\includegraphics[width=0.77\textwidth]{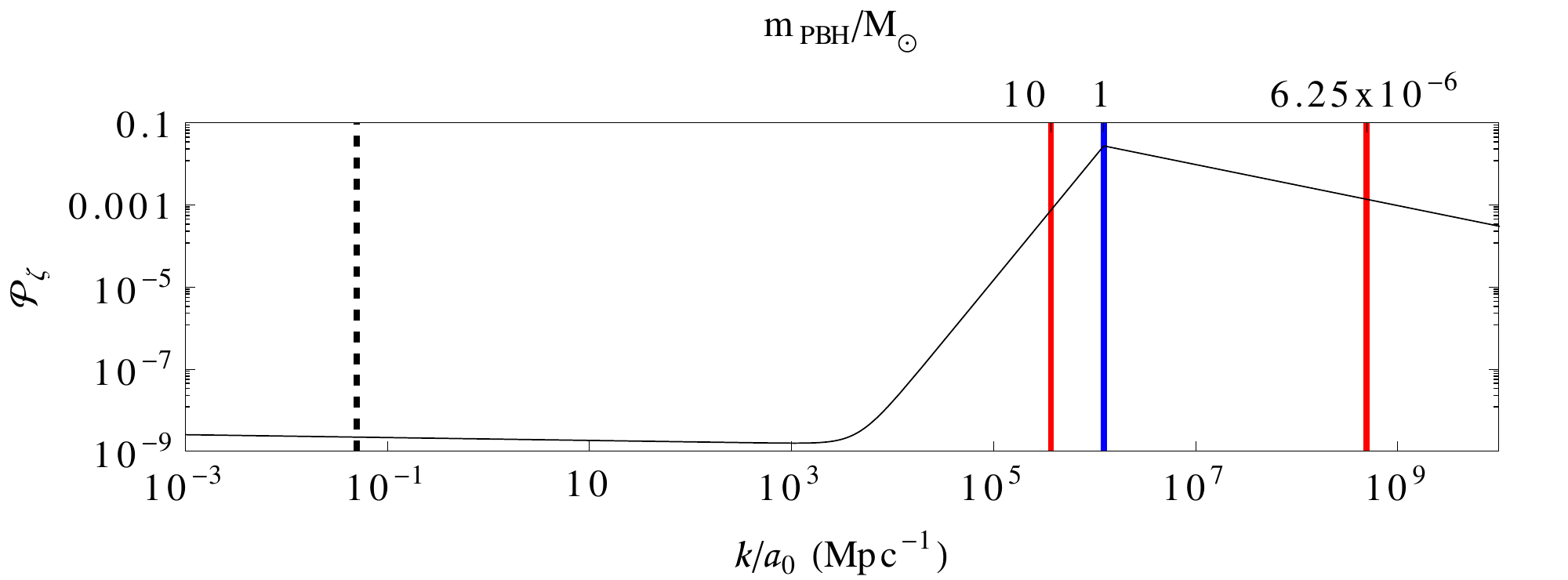}
\includegraphics[width=0.77\textwidth]{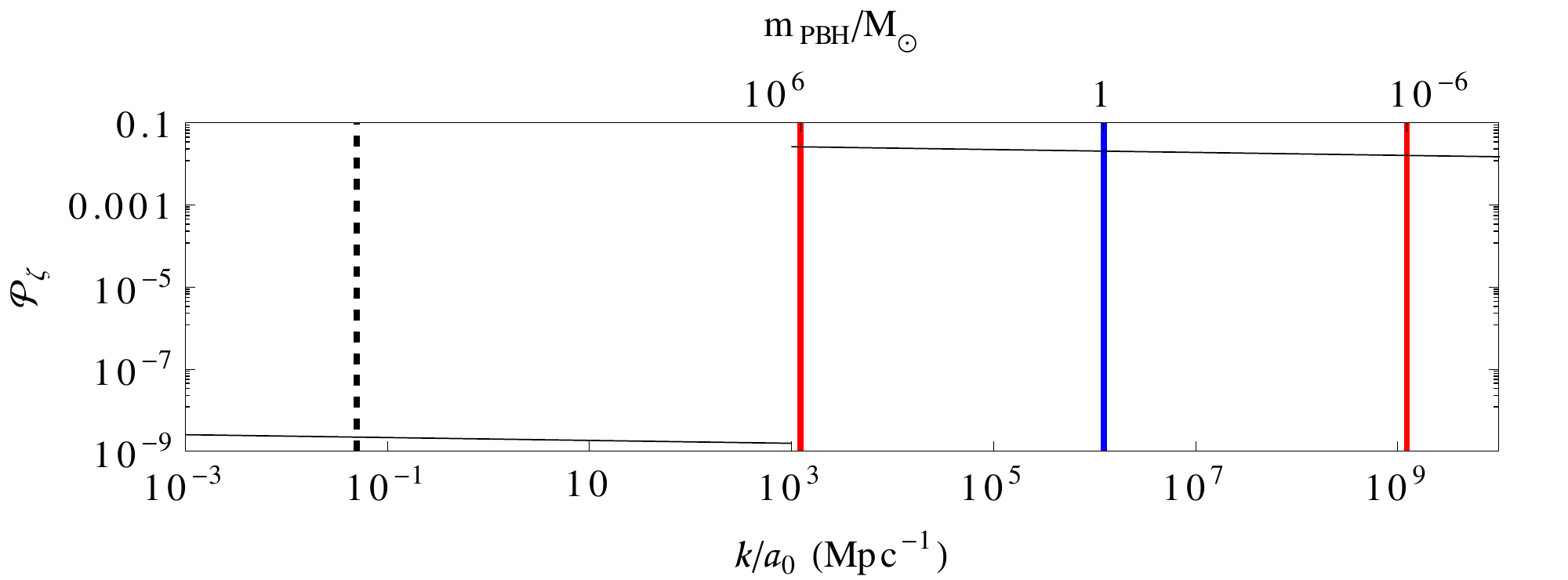}
\hfill
\caption{Spectrum of primordial curvature perturbations at the end of inflation $\mathcal{P}_{\zeta}$ as a function of $k$.  Top:  For the Gaussian model (\ref{eq:phi}), with $k_{\rm p} = 2 \times 10^6 {\rm Mpc}^{-1}$,  $\sigma_{\rm p } = 3$ and $\mathcal P_{\rm p} = 0.025 $.   Center-top:  For the sharp peak model, assuming a thin Gaussian with $\sigma_{\rm p} =  0.1$ and $\mathcal P _{\rm p}  = 0.0298 $.  Center-bottom:  For the broken power-law model (\ref{eq:pl}), with  $m=3$, $n=0.5$ and $\mathcal{P}_p=0.0275$.   Bottom:  for the nearly-flat model with boosted formation at QCD, with $n_{\rm p} = 0.96 $, $k_{\rm min} = 10^{3} {\rm Mpc}^{-1}$ and $P_{p} = 0.0205 $.  All models lead to $f_{\rm PBH}^{\rm tot} = 1$ today, i.e. PBH account for all the Dark Matter. The red vertical lines illustrate the window of modes considered for the evaluation of gravitational waves from PBH formation.}
\label{fig:specs}

\end{figure}

\begin{figure}[!tb] 
\centering
\includegraphics[width=0.8\textwidth]{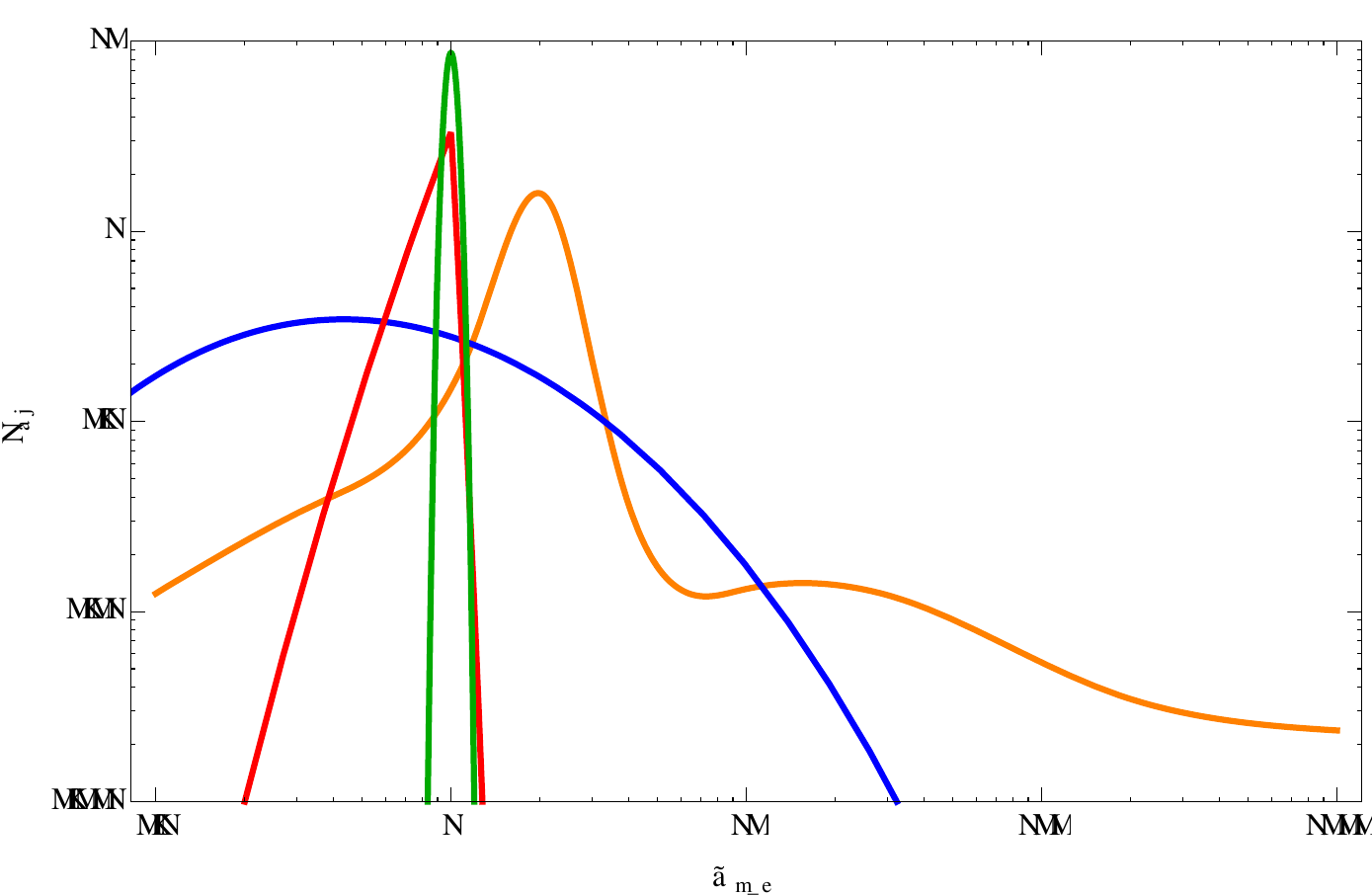}
\caption{PBH mass spectrum, $f_{\rm PBH} \equiv {\rm d } \rho_{\rm PBH} / {\rm d} \ln \mPBH$ vs. the PBH mass $\mPBH$ in solar mass units, for the considered models (blue: Gaussian peak, green: sharp peak, red: broken power-law, orange:  nearly-flat spectrum with QCD boost) and parameters as in Fig.~\ref{fig:specs}. All models lead to $f_{\rm PBH}^{\rm tot} = 1$, i.e. PBH make all the Dark Matter.}
\label{fig:massdist}

\end{figure}

In some models of bubble wall collisions and/or collapse after inflation~\cite{Deng:2016vzb,Deng:2017uwc,Cotner:2016cvr,Cotner:2017tir}, the power spectrum is a broken power law.
Here we consider a generic spectrum of primordial scalar perturbations. We will only assume that it has the functional form of a broken power law:
\be
\mathcal{P}_\zeta (k)\,=\, A_s\left(\frac{k}{k_*}\right)^{n_s-1} + \left\{
        \begin{array}{ll}
                \mathcal{P}_p \left(\frac{k}{k_{\rm p}}\right)^m & \;\; k<k_{p} \\
                \mathcal{P}_p \left(\frac{k}{k_{\rm p}}\right)^{-n} & \;\; k\geq k_{p} \\
        \end{array}
    \right.\,,
\label{eq:pl}
\en
where $m>0$ and $n>0$ are real numbers, $k_{\rm p}$ the peak scale and $\mathcal{P}_p<1$. An example of such a primordial spectrum, with $m=3$, $n=0.5$ and $\mathcal{P}_p=0.0275$ leading to $f_{\rm PBH}^{\rm tot} = 1$ today, is shown in Fig.~\ref{fig:specs}.

\subsection{Nearly flat power spectrum and boost of PBH formation at QCD transition}
\label{ssec:QCDpeak}

The sound speed reduction during the cross-over QCD phase transition reduces the threshold of the density contrast, and therefore boosts the formation of PBH~\cite{Jedamzik:1996mr,Cardall:1998ne}, when the Hubble-horizon mass is about $2 M_\odot$, as recently detailed in~\cite{Byrnes:2018clq}.   This produces a PBH mass function peaked on the sub-solar range, typically between $0.5$ and $2 M_\odot$ depending on the ratio between the  final PBH mass and the Hubble-horizon mass at re-entry.   In this case, there is no need of a peaked primordial power spectrum. A nearly scale-invariant spectrum, whose amplitude is nevertheless enhanced compared to CMB scales, can perfectly produce stellar-mass PBH with an abundance comparable to the DM, without overproducing much heavier or lighter PBH.   Such a scenario could be more natural than the one of a peak in the spectrum, since slow-roll inflation generically predicts a nearly scale-invariant power spectrum.   It was modeled like
\be
\mathcal{P}_\zeta (k)\,=\, A_s \left(\frac{k}{k_*}\right)^{n_s-1} + \mathcal{P}_p \left(\frac{k}{k_{\rm p}}\right)^{n_{\rm p}-1}  \Theta( k- k_{\rm min })~,
\en
where $\Theta (x) $ is the Heaviside function, $ k_{\rm min } $ denotes the mode at which the transition occurs, and $n_{\rm p}$ is the spectral index on PBH scales.  
 In order to compute the mass function, we followed~\cite{Byrnes:2018clq} and assumed for simplicity that the PBH mass corresponds to the mass of the Hubble volume at horizon re-entry of the curvature fluctuation.  But a different assumption would only shift the peak in the PBH distribution to lower masses and slightly change the relation between the PBH masses and GW frequencies.   We find that the spectral index $n_{\rm p} $ needs to take a value between $0.94$ and $0.98$, lower/larger values leading respectively to an overproduction of heavy/light PBH.   We considered as a benchmark $n_{\rm p} = 0.96 $ with $k_{\rm min} = 10^{3} {\rm Mpc}^{-1}$ and $\mathcal{P}_{\rm p} = 0.0205 $, which leads to $f_{\rm PBH}^{\rm tot} = 1$.   Since the details of the QCD transition are well-known and the speed of sound reduction is unavoidable, this is probably the best physically motivated model of PBH production.  

\section{Gravitational waves from the formation of primordial black holes}
\label{sec:gw}

During inflation, the quantum fluctuations in quasi-de Sitter expansion produce gravitational waves, and to first order in the perturbations of the Einstein equations, tensor fluctuations evolve independently of scalars. In slow-roll inflation models, these first-order gravitational waves turn out to be very weak and far below the sensitivity of any gravitational wave detector, but their imprint on the B-mode polarization of the CMB and could be observable by future dedicated instruments.  To second order, however, the scalar perturbations are a source of gravitational waves. If the power spectrum of scalar perturbations features a high peak, this second order effect can be the dominant one.
In \cite{Ananda:2006af,Baumann:2007zm}, general formulas are given to calculate the gravitational waves generated by any primordial scalar perturbations when they reenter the horizon after inflation. It is assumed that, immediately after a mode $k$ enters the horizon, the tensor perturbation $h_k$ grows and reaches a magnitude $S(t_k)/k^2$, where $t_k$ is the time when $k=a_k H_k$ and $S$ is the source term of gravitational waves. Then, with details depending on whether the mode enters the horizon during radiation or matter domination, the perturbation $h_k$ evolves to a final value, determining the spectrum of gravitational waves today. In practice, only very long-wavelength modes enter the horizon during matter domination. The modes of interest in our scenarios enter during radiation domination.

In what follows, we compute the spectrum of tensor perturbations $\mathcal{P}_{h}(k,t_k)$ at time $t_k$ and, via a transfer function $T(k,t)$, derive the spectrum today $\mathcal{P}_{h}(k,t_0)=T(k,t_0)\mathcal{P}_{h}(k,t_k)$ for the different primordial power spectra described in section \ref{sec:models}.
Here and below, a subscript $0$ indicates a quantity evaluated today and a subscript $k$ a quantity evaluated when the mode $k$ enters the horizon.

The starting point is to calculate the power spectrum $\mathcal{P}_{h}(k,t_k)$. It is given by eq.~(44) of \cite{Ananda:2006af,Baumann:2007zm}:

\be
\mathcal{P}_{h}(k,t_k)\,&\equiv&\,\frac{k^3}{2\pi^2}\braket{h_k(t_k)^2}\nn
                        &\simeq&\,\frac{1}{2\pi^2k}\int {\rm d}^3\mathbf{p}\ p^4(1-\mu^2)^2\mathbf{\Phi}^2(p/k)\mathbf{\Phi}^2(|\mathbf{k}-\mathbf{p}|/k)\frac{\mathcal{P}_\zeta(p)}{p^3}\frac{\mathcal{P}_\zeta(|\mathbf{k}-\mathbf{p}|)}{|\mathbf{k}-\mathbf{p}|^3}\,,
\label{eq:Ph}
\en
where $\mu\equiv \mathbf{k}\cdot\mathbf{p}/(kp)$ and the transfer function for first order scalar perturbations
\be 
\mathbf{\Phi}(p/k)\,&\simeq&\, \left\{
        \begin{array}{ll}
                (1+p^2/k^2)^{-1} & \;\; k \geq k_{\rm eq} \\
                (1+p^2/k_{\rm eq}^2)^{-1} & \;\; k < k_{\rm eq}\\
        \end{array}
    \right.\,.
\label{eq:bardeen}
\en 
We assumed that the Bardeen potentials are equal, $\mathbf{\Phi}=\mathbf{\Psi}$. In reality, anisotropic stresses due to the presence of photons and neutrinos can lead to a difference in the two potentials and neglecting them leads to errors of $\sim 10\%$ for both first and second order perturbations \cite{Ananda:2006af,Baumann:2007zm}. For our purposes, accounting for these errors is not necessary.   It is convenient to rewrite the three-dimensional integral of Eq.~\ref{eq:Ph} as an integral over $p$ and $\mu$.  This gives, for the modes of interest,
\be
\mathcal{P}_{h}(k,t_k)\, \simeq  \frac{1}{\pi k} \int_0^{\infty} {\rm d }p \int_{-1}^1 {\rm d} \mu \,  \frac{ p^3 \left(1- \mu^2 \right)^2 \mathcal P(p) \mathcal P(\sqrt{|k^2+p^2 - 2\mu k p|}) }{\left( 1 + \frac{p^2}{k^2 } \right)^2 \left( 1+ \frac{|k^2 + p^2 - 2 \mu k p|}{k^2} \right)^2 \left( |k^2+p^2 - 2\mu k p | \right)^{3/2}}~.
\en
For a given mode $k$ and assuming a scale invariant power spectrum, in the limit $p\ll k $, the integrant is suppressed like $\propto p^3 /k^3$.   In the opposite limit $p\gg k $, it is also suppressed, like $\propto k^8 / p^8$.  As a result, the modes giving the dominant contribution to the integral are the ones $ p \sim k$.   For a peaked power spectrum, the integrand is further suppressed in the regime far from $p \sim k_{\rm p} $, which induces a peak in the GW spectrum.

Then, in order to obtain the spectrum of gravitational waves today, we multiply Eq.~(\ref{eq:Ph}) by the appropriate transfer function (see \cite{Baumann:2007zm}),
\be 
T(k,t)\,&=&\, \left\{
        \begin{array}{ll}
                1 & \;\; k<k_{\rm eq} \\
                \left(\frac{k}{k_{\rm eq}}\right)^{-\gamma} & \;\; k_{\rm eq}<k<k_{\rm c} \\
                \frac{a_{\rm eq}k_{\rm eq}}{a(t)k} & \;\; k>k_{\rm c} \\
        \end{array}
    \right.\,.
\label{eq:omgw}
\en
where $k_{\rm c}(t) \equiv k_{\rm eq} [a(t)/a_{\rm eq}]^{1/(\gamma - 1)}$ is the critical wavenumber below which sub-horizon modes at time $t$ did not settle down.   The redshifting factor $\gamma$ is fixed by numerical simulations done in \cite{Baumann:2007zm} to $\gamma\sim3$.   The resulting relative energy density of gravitational waves induced by scalar perturbations today, $\Omega_{\rm GW,\, 0} $ is then given by
\be
\Omega_{\rm GW,\,0}\,&=&\,\frac{a_0 k^2}{a_{\rm eq}k_{\rm eq}^2}T^2(k,t_0)\mathcal{P}_h (k)\,.
\en
In the subsections below, we use these results to calculate the spectra of GW produced at second order in the models described in Sections \ref{ssec:hi}, \ref{ssec:bpl}, \ref{ssec:delta} and \ref{ssec:QCDpeak}.

\subsection{GW from Gaussian power spectrum}
\label{ssec:gwhi}

%As explained in Section \ref{ssec:hi}, the primordial spectrum $\mathcal{P}_\zeta$ of eq.~(\ref{eq:phi}) is invariant when the combination of parameters $\Pi=M\sqrt{\mu_1\phi_c}$ is fixed. Therefore, it is sufficient to calculate eq.~(\ref{eq:Ph}) for fixed values of $\Pi^2\lesssim 200$. To this end, we will use the analytical expression of eq.~(\ref{eq:phi}), which, as can be seen in fig.~\ref{fig:specs} provides a very good fit to the spectrum of primordial perturbations. It is important to keep in mind that, although $\mathcal{P}_\zeta$ only depends on $\Pi$, different values of $\Lambda$ lead to different frequencies in the gravitational waves' spectrum. This can be seen clearly in eq.~(\ref{eq:Nstar}): the amount of inflation between horizon exit of a given scale $k$ and its re-entry depends on the energy scale at the end of inflation. For a fixed $\Pi$, the transformation $\Lambda\rightarrow\Lambda^{'}$ implies $\nu\rightarrow\nu^{'}=\nu(\Lambda^{'}/\Lambda)^{1/4}$. 

\begin{figure}[!tb]
\centering
\includegraphics[width=0.85\textwidth]{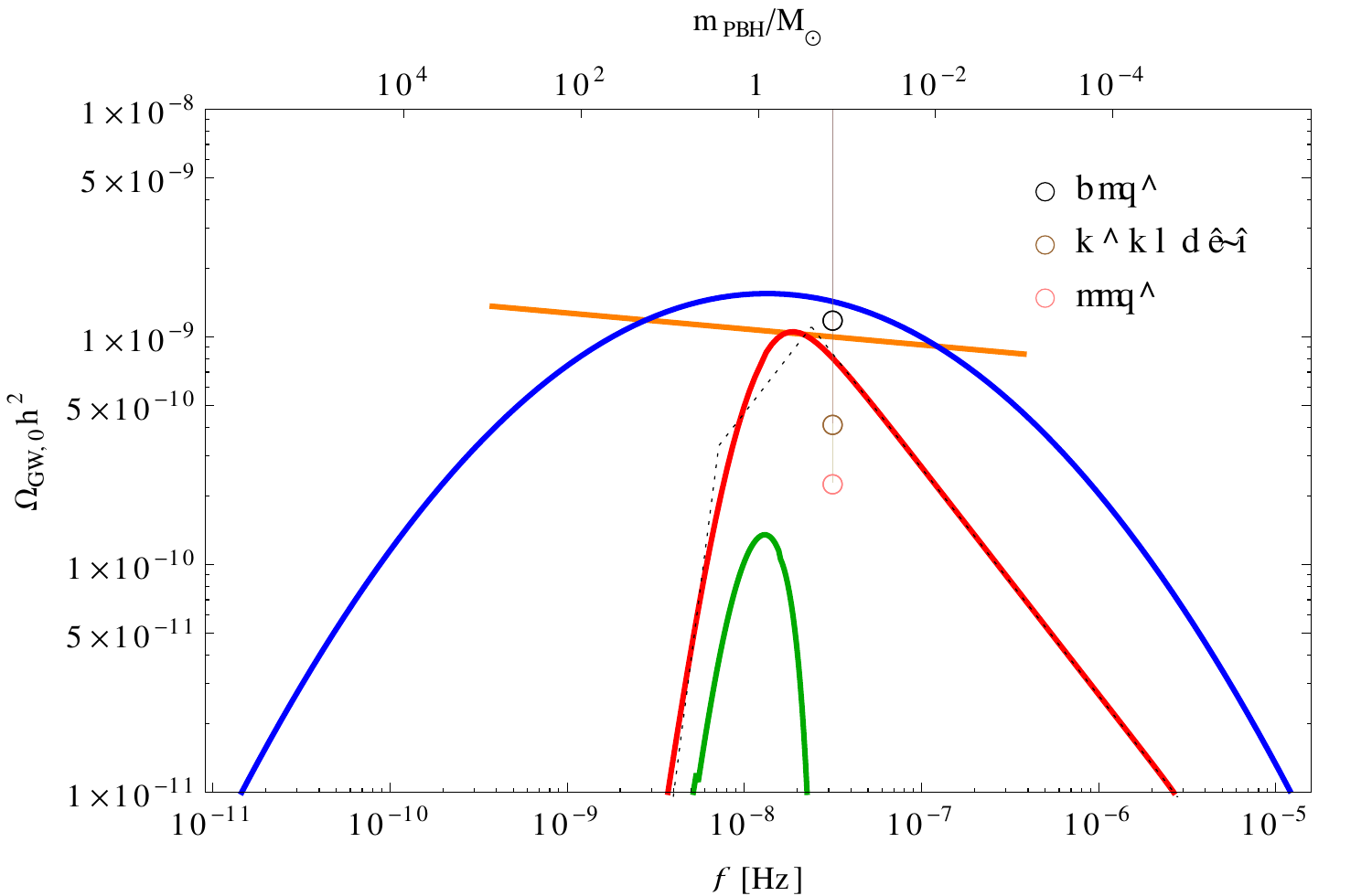}
\hfill
\caption{Spectra of the stochastic gravitational wave background from PBH formation, for the models of Fig.~\ref{fig:specs}.  The circles show the best current 95\% C.L. limits from different pulsar timing arrays (PPTA~\cite{Lasky:2015lej}, EPTA~\cite{Lentati:2015qwp} and NANOGrav~\cite{Arzoumanian:2015liz}), at frequency $f=1\, {\rm yr}^{-1}$ (assuming scale-invariant GW spectrum).  }
\label{fig:gwfinal}
\end{figure}

We have evaluated numerically the integral of Eq.~(\ref{eq:Ph}) for one hundred $k$ values in the range $[k_{\rm p}\times10^{-3}, k_{\rm p}\times10^{3}]$, where $k_{\rm p}$ is the peak scale of $\mathcal{P}_\zeta$.   The resulting gravitational wave spectrum is shown in Fig.~\ref{fig:gwfinal}.
As can be seen in Fig.~\ref{fig:specs}, this range of scales is sufficient to capture practically all the power of the perturbations. We checked that the result is stable under changes in the number of modes and domain of integration.   The GW spectrum amplitude peaks at  $\Omega_{\rm GW, 0} h^2 \approx 1.5 \times 10^{-9}$ on the nanohertz scale probed by PTA.   We also computed the spectrum of GW from PBH formation corresponding to a different peak scale $k_{\rm p}$ and adapted the peak amplitude $\mathcal P_{\rm p} $ to still get $f_{\rm PBH}^{\rm tot}= 1$ today, so that the GW spectrum peaks  in the range of LISA with PBH mass $\mPBH \sim 10^{-10}\ M_\odot$.   This spectrum is shown in Fig.~\ref{fig:gwfinal2}.

\subsection{GW from a sharp peak in the power spectrum}
\label{ssec:gwdelta}

%\begin{figure}[!tb]
%\centering
%\includegraphics[width=0.85\textwidth]{graphics/GWdelta.pdf}
%\hfill
%\caption{Spectrum of tensor perturbations sourced by (\ref{eq:delta}) with $\mathcal{P}_p=0.1$ as obtained from eq.~\ref{eq:gwm}. The vertical dashed line corresponds to the scale $k_{\rm p}=1.23\times10^6 Mpc^{-1}$. It is striking that the power of the GW spectrum is much lower than figs.~\ref{fig:gw} and \ref{fig:gwbr}. This is due to the $\delta$-function spikyness, implying that only one mode contributes to the formation of PBHs and production of GW.}
%\label{fig:gwdelta}
%\end{figure}

The gravitational waves from a single scalar mode spectrum, as in Eq.~(\ref{eq:delta}), can be calculated analytically as follows:
\be
%\mathcal{P}_{h}(k,t_k) \,&\simeq&\, \frac{\mathcal{P}_p^2}{\pi k} \int_0^{\infty} {\rm d} p \int_{-1}^1 \dd \mu\ p^3 (1-\mu^2)^2\mathbf{\Phi}^2(p/k)\mathbf{\Phi}^2(|\mathbf{k}-\mathbf{p}|/k)\frac{\delta(p-k_{\rm p})\delta(|\mathbf{k}-\mathbf{p}|-k_{\rm p})}{|\mathbf{k}-\mathbf{p}|^3} \nn 
%&=&\,\frac{\mathcal{P}_p^2}{16\pi}\frac{k^6(k^2-4k_{\rm p}^2)^2}{k_{\rm p}^4(k^2+k_{\rm p}^2)^4}\Theta(2k_{\rm p}-k)\,,
\mathcal{P}_{h}(k,t_k) \,&\simeq&\, \frac{\mathcal{P}_{\rm p}^2}{\pi k} \int_0^{\infty} {\rm d} p \int_{-1}^1 \dd \mu\ p^3 (1-\mu^2)^2\mathbf{\Phi}^2(p/k)\mathbf{\Phi}^2(|\mathbf{k}-\mathbf{p}|/k)\frac{\delta(p-k_{\rm p})\delta(|\mathbf{k}-\mathbf{p}|-k_{\rm p})}{|\mathbf{k}-\mathbf{p}|^3} \nn 
&=&\,\frac{\mathcal{P}_{\rm p}^2 }{16\pi  }\frac{k^7(k^2-4k_{\rm p}^2)^2}{k_{\rm p}^3(k^2+k_{\rm p}^2)^4}\Theta(2k_{\rm p}-k)~,
% \simeq \,\frac{\mathcal{P}_{\rm p}^2 k_{\rm p}}{16\pi k }\frac{k_{\rm eq}^7(k^2-4k_{\rm p}^2)^2}{k_{\rm p}^{11}}\Theta(2k_{\rm p}-k)\,,
\label{eq:gwm}
\en
where integrating over the $\delta$-functions implies $p=k_{\rm p}$ and $\mu=k/(2k_{\rm p})$. The Heaviside function $\Theta$ is introduced because a perturbation with wavenumber $k_{\rm p}$ cannot generate at second order gravitational waves with wavenumber $k>2k_{\rm p}$. It is interesting to note that for $k\ll k_{\rm p}$, $\mathcal{P}_{h}(k,t_k) \propto k^7/k_{\rm p}^3$.   It ends abruptly at $k = 2 k_{\rm p}$.   For a Gaussian narrow peak, the transition is smoothed.    
In Fig.~\ref{fig:gwfinal} we show the resulting spectrum of gravitational waves, computed numerically for a sharp Gaussian peak with $\sigma_{\rm p} = 0.1$.   The analytical calculation reproduces well the slope of the GW spectrum at $k < k_{\rm p}$, whereas the slope at $k \gtrsim k_{\rm p}$ is due to the Gaussian power spectrum suppression.  Furthermore one can see that the maximal GW spectrum amplitude is about one order of magnitude lower than for models (\ref{eq:phi}) and (\ref{eq:pl}), despite the fact that the spectrum peak $\mathcal P_{\rm p} $ is slightly larger to keep $f_{\rm PBH}^{\rm tot} = 1$ constant.  The GW spectrum reaches $\Omega_{\rm GW, 0} h^2 \approx 1.5 \times 10^{-10}$.   This suppression compared to the cases of broad peaks in the primordial power spectrum, is due i) to the suppressed contribution of long wavelength modes $k \ll k_{\rm p}$ in the integral, ii) to the fact that modes larger than $2 k_{\rm p}$ do not contribute to the integral thanks to the Heaviside function.   But contrary to a naive expectation, the GW spectrum is not as sharp as the power spectrum peak.

%that PBH are formed only when the scale $k_{\rm p}$ reenters the horizon, and only then GW are produced, as opposed to the other models where a range of scales contributes to the production of GWs. 

\subsection{GW from a broken power law}
\label{ssec:gwpl}

%\begin{figure}[!tb]
%\centering
%\includegraphics[width=0.85\textwidth]{graphics/GWbr.pdf}
%\hfill
%\caption{Spectrum of gravitational waves today, $\Omega_{\rm GW,\,0}$, versus frequency  $\nu=k/a_0\,9.71\times10^{-15}\,{\rm Mpc\,Hz}$, for model (\ref{eq:pl}) with with $m=3$, $n=0.5$ and $\mathcal{P}_p=0.1$. The peak scale is shifted to illustrate the GW spectra corresponding to different PBH masses $M_{\rm PBH}$. The dotted lines are obtained from a numerical integration of eq.~(\ref{eq:Ph}), whereas the black line uses the analytical approximation of eq.~(\ref{eq:Phplfit}). The numerical factors $\mathcal{N}$, $\mathcal{N_I}$ and $\mathcal{N_{II}}$ are determined by comparison to the numerical solution. The agreement is very good.}
%\label{fig:gwbr}
%\end{figure}

To calculate the gravitational waves sourced at second order by the broken power law spectrum of Eq.~(\ref{eq:pl}), we can solve Eq.~(\ref{eq:Ph}) analytically. As already mentioned, we can safely consider $k<k_{\rm eq}$ for all the wave-vectors of interest here. We can separate the integral in three different parts: the region $k\ll k_{\rm p}$, the region $k\simeq k_{\rm p}$ and the region $k\gg k_{\rm p}$. 

Let's start by considering $k\ll k_{\rm p}$. The most relevant contribution in this region comes from the convolution with $p\ll k_{\rm p}$. Inserting the spectrum of Eq.~(\ref{eq:pl}) into Eq.~(\ref{eq:Ph}), one finds
\be
\mathcal{P}_{h}(k \ll k_{\rm p},t_k) \,&\simeq&\, \frac{2\mathcal{P}_{\rm p}^2}{\pi} \left(\frac{k}{k_{\rm p}}\right)^{2m} \int_0^{k_{\rm p}/k} \dd x \int_{\mu_{\rm p}}^1 \dd\mu \frac{(1-\mu^2)^2(1+x^2-2x\mu)^{\frac{m-3}{2}}x^{m+3}}{(2+x^2-2x\mu)^2(1+x^2)^2}\nn
&=&\, \mathcal{P}_{\rm p}^2 \left(\frac{k}{k_{\rm p}}\right)^{2m} \mathcal{N_I} \,
\label{eq:gwm2}
\en
where we defined the variable $x\equiv p/k$. The upper limit of the integral over $x$ goes to infinity for $k \ll k_{\rm p}$, whereas the lower limit of the integral over $\mu$ goes to $-1$. For this reason, the result of the integral will be just a number $\mathcal{N_I}$.
  
Similarly, for $k\gg k_{\rm p}$, the most relevant contribution comes from $p\gg k_{\rm p}$:
\be
\mathcal{P}_{h}(k \gg k_{\rm p},t_k) \,&\simeq&\, \frac{2\mathcal{P}_p^2}{\pi} \left(\frac{k}{k_{\rm p}}\right)^{-2n} \int_{k_{\rm p}/k}^{\infty} \dd x \int^{\mu_{\rm p}}_{-1} \dd \mu \frac{(1-\mu^2)^2(1+x^2-2x\mu)^{\frac{m-3}{2}}x^{m+3}}{(2+x^2-2x\mu)^2(1+x^2)^2}\nn
&=&\, \mathcal{P}_p^2 \left(\frac{k}{k_{\rm p}}\right)^{-2n} \mathcal{N_{II}} \,.
\label{eq:gwm3}
\en

Finally, the region $k\simeq k_{\rm p}$. Here the domains of integration are more complicated but, by analogy to the other regions, we can guess that around $k_{\rm p}$, $\mathcal{P}_{h}(k \simeq k_{\rm p},t_k) \propto k^{m-n}$. Putting together these results,  one has
\be
\mathcal{P}_h (k,t_k)\,=\, \left\{
        \begin{array}{ll}
                \mathcal{P}_p^2\mathcal{N_I}\left(\frac{k}{k_{\rm p}}\right)^{2m} & \;\; k\lesssim k_{p}\\
                \mathcal{P}_p^2\mathcal{N}\left(\frac{k}{k_{\rm p}}\right)^{m-n} & \;\; k\simeq k_{p}\\                
\mathcal{P}_p^2\mathcal{N_{II}}\left(\frac{k}{k_{\rm p}}\right)^{-2n} & \;\; k\gtrsim k_{p} \\
        \end{array}
    \right.\,.
\label{eq:Phplfit}
\en
In order to check the accuracy of these analytical approximations, in Fig.~\ref{fig:gwfinal} we compare them to a numerical calculation, for the spectrum of Fig.~\ref{fig:specs}. One can see that Eq.~(\ref{eq:Phplfit}) roughly captures the $k$ dependence of the GW spectrum.   Nevertheless, the numerical integration should be used for more accuracy and is nevertheless needed to compute the numerical factors $\mathcal N_{\mathcal I}$ and $\mathcal N_{\mathcal {II}}$.   For the considered model parameters, we find that the power spectrum peaks on nanohertz frequencies at $\Omega_{\rm GW, 0} h^2 \approx 1 \times 10^{-9}$.

\subsection{GW from nearly flat power spectrum and QCD transition boost}
\label{ssec:gwQCD}

Similarly to the broken power-law model, the gravitational wave spectrum can be calculated analytically as 
\be
\mathcal{P}_{h}(k ,t_k) \,&\simeq&\, \frac{2\mathcal{P}_{\rm p}^2}{\pi} \left(\frac{k}{k_{\rm p}}\right)^{2 (n_{\rm p} -1)} \int_0^{\infty} dx \int_{\mu_{\rm p}}^1 d\mu \frac{(1-\mu^2)^2(1+x^2-2x\mu)^{\frac{m-3}{2}}x^{m+3}}{(2+x^2-2x\mu)^2(1+x^2)^2}\nn
&=&\, \mathcal{P}_{\rm p}^2 \left(\frac{k}{k_{\rm p}}\right)^{2 (n_{\rm p} -1)}\mathcal{N_{III}} \,
\label{eq:gwm}
\en
but now the integral over $x$ encompasses both modes larger and lower than $k_{\rm p}$.   The resulting k-dependence is in agreement with the numerical computation of the gravitational wave power spectrum, which is shown in Fig.~\ref{fig:gwfinal}.    As a result, the stochastic GW spectrum on the nanoHertz scale is also nearly scale invariant, with an amplitude  $\Omega_{\rm GW, 0} h^2 \approx 1 \times 10^{-9}$.    But since there is no power spectrum peak, the GW spectrum could extend up to much larger frequencies, eventually covering the frequency range of space and Earth-based GW interferometers, with interesting perspectives of detection.
Discussing the current limits and prospects of detection for the different models is the goal of the next section.   

\section{Current limits and prospects of detection with PTA and GW interferometers}
\label{sec:obs}

%\begin{figure}[!tb]
%\centering
%\includegraphics[width=0.85\textwidth]{graphics/GWexperiments.pdf}
%\hfill
%\caption{}
%\label{fig:gwex}
%\end{figure}

The amplitude of the stochastic GW background from PBH formation is related to the power spectrum of primordial curvature fluctuations, whereas the GW frequency is linked to the PBH mass.   The mass scale relevant for black hole mergers detected by GW interferometers, i.e. $[1-50] \Msun$, interestingly coincides with the frequency range covered by PTA experiments, i.e. the nanoHertz scale.   PTA are therefore the natural and ideal probe to detect or constrain the existence of stellar-mass PBH, and to distinguish between different formation scenarios. 

The first important result of this paper is that for the considered models producing a wide mass distribution of PBH, the stochastic GW spectrum amplitude is enhanced compared to the simpler but less realistic monochromatic (or close to monochromatic) mass model that was considered so far.   This enhancement is about one order of magnitude, so that $\Omega_{\rm{GW,0}} (f_{\rm peak}) h^2 \sim 10^{-9} $.  This makes the considered models and parameters already in strong tension with the best and current PTA limits from the NANOGrav~\cite{Arzoumanian:2015liz} and the Parkes Pulsar Timing Array (PPTA)~\cite{Lasky:2015lej} collaborations, as shown on Fig.~\ref{fig:gwfinal}, whereas the limits set by the European Timing Arrays (EPTA)~\cite{Lentati:2015qwp} can still be accommodated.   Note that we considered the limits at $f= \rm yr^{-1}$ set for a scale-invariant spectrum, but they should be relatively accurate for our GW spectra peaking at frequencies where PTA are the most sensitive, making them roughly scale-invariant in this frequency range.  
%(see e.g. Fig.? of~\cite{} for constraints with different power-law indexes).  

It is however important to remember, on the one hand, that the peak amplitude in the GW background spectrum goes like $\Omega_{\rm{GW,0}} (f_{\rm peak}) \propto \mathcal P_{\rm p}^2$, and on the other hand, that the PBH abundance is very sensitive to power spectrum peak amplitude and to the critical curvature threshold, going roughly like $\beta \propto (\zeta_c / \sqrt{\mathcal P_{\rm p}}) \exp (- \zeta_{\rm c}^2 / 2 \mathcal P_{\rm p}) $.   Therefore the most crucial parameter combination is actually $\zeta_{\rm c} / \sqrt{\mathcal P_{\rm p}}$.  At fixed abundance of PBH, a lower value of $\zeta_{\rm c} $ must be compensated by a lower value of $\mathcal P_{\rm p}$ to keep their ratio constant.   For instance, if one assumes $\zeta_{\rm c} =0.086$ (instead of $\zeta_{\rm c} =1.02$), as obtained for a semi-numerical three-zone model of PBH formation~\cite{Harada:2013epa}, one needs $\mathcal P_{\rm p} $ to be about 150 times lower than the values considered in Fig.~\ref{fig:specs}.   In turn, the peak amplitude in the GW spectrum is strongly suppressed, by a factor $ 2 \times 10^4$, down to $\Omega_{\rm{GW,0}} (f_{\rm peak}) h^2 \sim 10^{-13}$, which makes it compatible with all the present PTA limits.  That roughly gives the plausible and relatively wide range for the GW background amplitude associated to PBH formation.  It will be partially covered by the 5-year observations from the International Pulsar Timing Array (IPTA)~\cite{2016MNRAS.458.1267V} that has a projected sensitivity of $\Omega_{\rm{GW,0}} h^2 \sim 10^{-10} $~\cite{Lasky:2015lej}, and in a farther future by the PTA from the Square Kilometre Array whose sensitivity could reach $\Omega_{\rm{GW,0}} h^2\sim 10^{-15} $~\cite{Lazio:2013mea,Zhao:2013bba}.  These considerations are  valid not only for the three wide-mass distribution models considered in the paper, but they are generic to any scenario producing a wide PBH mass distribution from Gaussian curvature fluctuations.  As a consequence, if the GW spectrum from PBH formation is detected with PTA, its shape will provide an important information about the PBH formation process, and thus on the underlying inflation scenario.   If there is no detection of a GW background by IPTA and SKA, then even a tiny fraction of PBH will be ruled out.  The reason is that one roughly has  $\Omega_{\rm{GW,0}}  \propto (\ln \beta_{\rm form})^{-2}$, so that ultimately, with the SKA sensitivity, one would probe values of $\mathcal P_{\rm p} < 10^{-4} $ and one would be able to rule out the existence of even a single stellar-mass PBH in our Observable Universe.   A noticeable exception are models where PBH originate from highly non-Gaussian fluctuations~\cite{Garcia-Bellido:2017aan}. In such a case, the typical amplitude of curvature fluctuation can be much lower than assumed here, as recently studied in~\cite{Nakama:2016gzw,Franciolini:2018vbk,Ezquiaga:2018gbw,Byrnes:2012yx}.

\begin{figure}[!tb]
\centering
\includegraphics[width=0.85\textwidth]{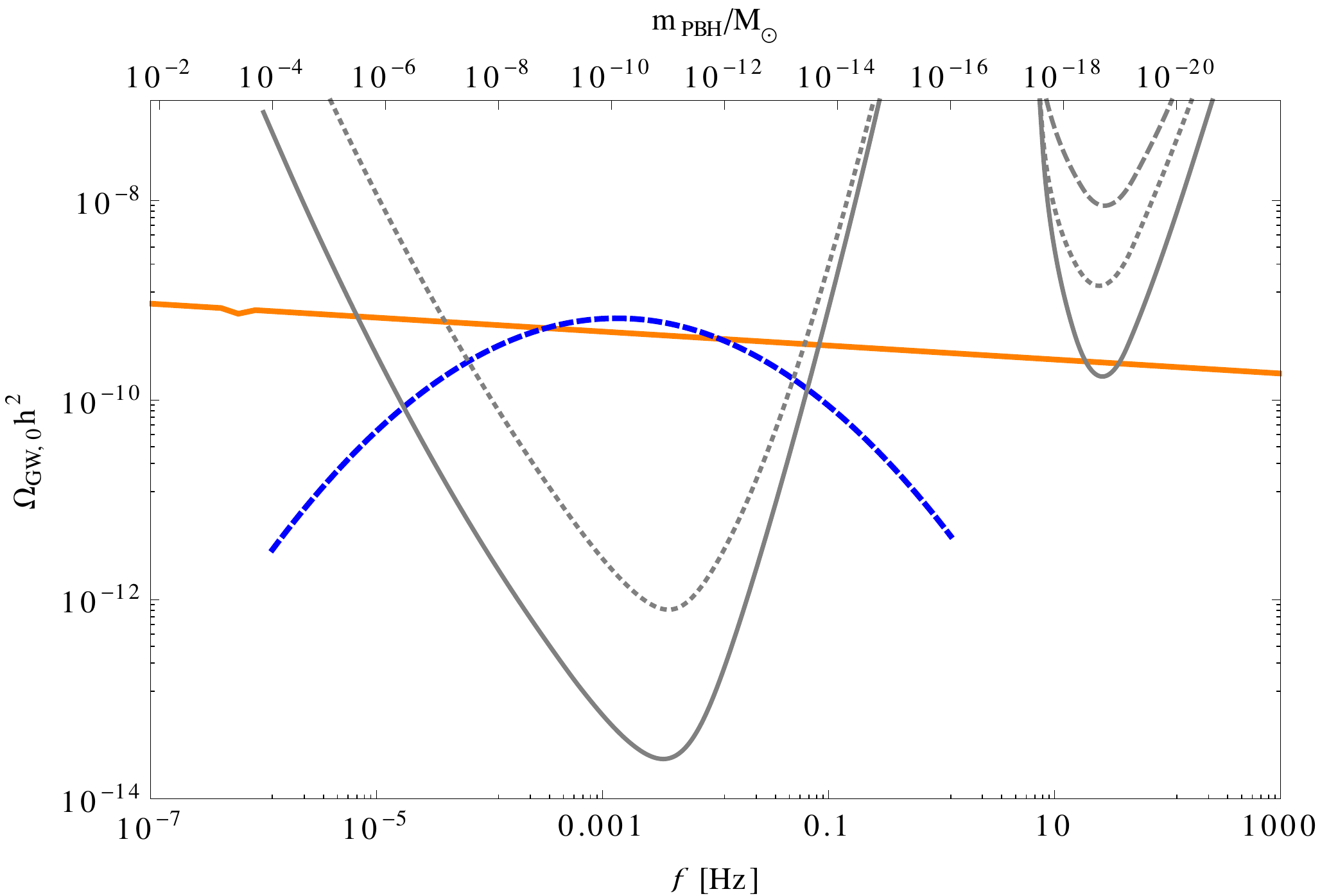}
\hfill
\caption{GW background spectrum from PBH formation, for a gaussian peak centered on $\mPBH = 10^{-10} M_\odot$ with $\mathcal P_{\rm p} =0.0163 $ and $\sigma_{\rm p} = 3$ so that $f_{\rm PBH} = 1$ today (dashed blue line), and for a nearly scale-invariant power spectrum with $n_{\rm p} = 0.96$ and boosted PBH formation at QCD transition (orange line), as in Fig.~\ref{fig:gwfinal} but extended to higher frequencies.   Gray lines represent the projected sensitivity of LISA (left), for the best (solid) and worse (dotted) experimental designs, and the limits or expected sensitivity of LIGO/Virgo (right - O1 Run - dashed, O2 Run - dotted, O5 Run - solid). }
\label{fig:gwfinal2}
\end{figure}

In some models such as with a broken power-law spectrum with a second index $n \lesssim 0.3$, or with a wide Gaussian peak with $\sigma_{\rm p} \gtrsim 4$, the GW background spectrum extends to larger frequencies that are relevant for LISA.   Space interferometers will therefore be excellent complementary probes to discriminate between different PBH models, by increasing the frequency lever arm.    LISA will also constrain PBH models with lower mean masses, down to $\mPBH \sim 10^{-14} \Msun$.  One example of GW spectrum from a Gaussian peak whose parameters are adapted to get $f_{\rm PBH}^{\rm tot} = 1$ is shown in Fig.~\ref{fig:gwfinal2}.  One can see that even if $\beta_{\rm form}$ must be strongly suppressed compared to a stellar-mass PBH model with the same abundance today, the peak amplitude in the GW background only changes by a factor of a few.  This behavior is universal and comes once again from the exponential dependence of the PBH abundance to the peak amplitude.    This is an interesting mass range because it is not well constrained by astrophysical probes (note, however, constraints from white dwarfs and neutron stars in globular clusters~\cite{Capela:2013yf,Defillon:2014wla,Capela:2012jz,Pani:2014rca} whose strength is debated), which allows an important fraction of Dark Matter to be composed by such light PBH. The Space mission LISA should have the sensitivity to definitively detect or rule out these models~\cite{Bartolo:2016ami}.  

On frequencies probed by Earth-based GW interferometers like LIGO/Virgo and the future Einstein Telescope, KAGRA and LIGO-India, the GW spectrum from PBH formation would be induced by extremely light PBH, with $\mPBH \lesssim 10^{12} {\rm kg}$, light enough to evaporate in a time much shorter than the age of the Universe, but longer than the time of Big-Bang nucleosynthesis.   Their abundance is thus strongly constrained and such PBH could not significantly contribute to the Dark Matter today.  Nevertheless, in the case of the model with a scale-invariant power spectrum and a boost of PBH formation at the QCD transition, it is likely that the power spectrum extends up to the frequencies of interest, if the second slow-roll phase responsible for the power spectrum enhancement continues until the end of inflation.   In this case, we find that the GW spectrum from the formation of light but subdominant PBH, shown in Fig.~\ref{fig:gwfinal2}, passes the current limits imposed by LIGO/Virgo but could be detected with upcoming runs, as well as with future GW detectors.   Ground-based interferometers will therefore be an additional mean to probe the existence of PBH for this particular model, that is actually the best theoretically motivated.

\section{Conclusions}\label{sec:conclusions}

The renewed interest for stellar-mass PBH since the detection of gravitational waves from black hole mergers motivates us to study the stochastic gravitational wave background that is produced at the same time of their formation.  The GW production mechanism relies on the large level of scalar perturbations which, at second order, are a source of tensor modes.  These PBH could eventually constitute an important fraction, or even the totality of the Dark Matter in the Universe.   
Compared to previous work, we did not only consider  the expected GW background for a monochromatic PBH mass model, but also for different wide-mass distributions that are typical of theoretical scenarios of PBH formation, such as mild-waterfall hybrid inflation, axion inflation, or boosted PBH formation during the QCD cross-over transition.  In the latter case, a nearly scale invariant power spectrum of curvature fluctuations, but enhanced with respect to CMB scales, can produce a dominant population of PBH at the solar-mass scale.   Three other models were considered:  a wide Gaussian peak, a broken power-law and a thin peak in the curvature power spectrum.   Those wide-mass models could explain the LIGO/Virgo black hole properties and merger rates, while escaping the astrophysical constraints on PBH abundance if one takes into account that micro-lensing constraints are  questioned and can be evaded when considering realistic dark matter profiles in our Galaxy or the possible clustering of PBH.   The formation of stellar-mass black holes typically produces a nanohertz gravitational wave background, an ideal target for Pulsar Timing Arrays.

For these models and for typical sets of parameters, we have evaluated numerically the frequency spectrum of the stochastic GW background, and for some of them we derived some analytical approximations. The first result of this paper is that the amplitude of the stochastic gravitational wave background for wide-mass models is enhanced by about one order of magnitude compared to the monochromatic case, because of an enhanced contribution of different wavelength modes around the peak scale to a given GW frequency.    Assuming a critical curvature fluctuation threshold for PBH formation of order unity, as predicted by simulations in numerical relativity~\cite{Musco:2004ak,Shibata:1999zs}, this makes wide-mass PBH models already in strong tension with the best limits set by the NANOGrav and Parkes Pulsar Timing Arrays.   However, a lower threshold value, e.g. the one predicted by the three-zone semi-numerical model of Harada et al.~\cite{Harada:2013epa} implies a lower GW background that can easily accommodate the current observational limits.   Models involving non-Gaussian fluctuations would also generate a suppressed GW background~\cite{Nakama:2016gzw}.  

We also found that the GW background amplitude is rather insensitive to the mean PBH mass and to the fraction of dark matter made of PBH, because the latter is exponentially dependent of the primordial power spectrum of curvature fluctuations.   For any mean PBH mass scale, the stochastic background from broad-peak PBH formation will have an amplitude of $\Omega_{\rm{GW,0}} h^2 \sim 10^{-9} $ for the largest plausible value of the critical threshold, and down to $\Omega_{\rm{GW,0}} h^2 \sim 10^{-13} $ for the smallest one.   As a consequence, LISA will probe the existence of light PBH, down to masses of $10^{-14} M_\odot$.   Another consequence is that for a nearly scale-invariant power spectrum with a naturally boosted PBH formation at the QCD transition, LISA but also Earth-based GW interferometers such as LIGO/Virgo and the future KAGRA, LIGO-India and Einstein Telescope will have the sensitivity to detect of the stochastic GW background from the formation of light PBH that is predicted in this best theoretically motivated scenario.  

Finally, we have emphasized that future PTA with the Square Kilometre Array could exclude the existence of even a single PBH in our observable Universe (again, assuming they are formed from Gaussian fluctuations).  Searching for and setting limits on the possible stochastic GW background induced by large scalar fluctuations is therefore a way to set much more stringent limits ont the existence of PBH than any other astrophysical mean, and to distinguish between different formation scenarios.   This should therefore be considered as an important task for the the coming years.   In order to reach this objective, theoretical progresses will be also needed, such as a better determination of the critical curvature threshold for different realistic PBH formation scenarios, including the effect of possible non-Gaussianity, and a more accurate computation of the stochastic background amplitude, e.g. by using more accurate transfer functions.

\section*{Acknowledgements}
We thank M. Peloso and C. Byrnes for useful comments and discussions.  JGB is supported by the Research Project FPA2015-68048-C3-3-P [MINECO-FEDER], and the Centro de Excelencia Severo Ochoa Program SEV-2016-0597. JGB thanks the Theory Department at CERN for their hospitality during a Sabbatical year at CERN. He also acknowledges support from the Salvador de Madariaga Program Ref. PRX17/00056.
SO was supported by the Swiss National Science Foundation.   The work of SC is supported by a \textit{Charg\'e de Recherche} grant of the Belgian Fund for Research F.R.S.-FNRS.

\end{document}